# Measurement Of The Fluid Flow Load On A Globe Valve Stem Under Various Cavitation Conditions

**Presenter:**

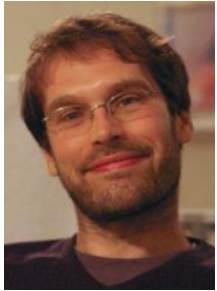

### Jerome FERRARI
jerome.ferrari@edf.fr
  2002 - *Licentiate in applied fluid mechanics*, Royal Institute of Technology, Stockholm, Sweden.
  1997 – *Master degree in mechanical engineering*, Ecole Nationale Supérieure de Mécanique et Energétique de Valenciennes.
Present professional position:
Research engineer at EDF R&D in the Material and Mechanics of Components Department, Centre des Renardières, Moret-sur-Loing, France. Working on fluid mechanics aspects of nuclear valves since 2004.

### Zachary Leutwyler
zleutwyler@kalsi.com
  2004 – MSME, University of Houston, Houston, TX
  2002 – BSME, University of Houston, Houston, TX
Present professional position:
Consultant Engineer at Kalsi Engineering, Inc.
Areas of research include experimental and numerical investigation of flow-induced forces and torque acting on linear and quarter turn valves.

# Measurement Of The Fluid Flow Load On A Globe Valve Stem Under Various Cavitation Conditions

Jérôme Ferrari, EDF R&D, Moret-sur-Loing
Zachary Leutwyler, Kalsi Engineering

## Abstract:
The evaluation of fluid forces on the stem is important for wear prediction of valve and actuator guidance parts. While estimating the axial load is straight forward, estimating and/or measuring the side load is more difficult, especially for globe valves. Therefore, measurements are carried out on an ad hoc, scale 1, model of a 2" globe valve. The body is replicated in Plexiglas to enable flow visualization and the original stem is heavily instrumented to allow force measurements in every direction. The flow loop facility used for this experiment is designed to allow fine-tuning of the cavitation intensity.

The experiment is run for a set of cavitation conditions, flow rates and disc positions. The results show that the transverse force (perpendicular to the stem) can reach the order of the axial force; thus it should not be ignored. We also observed that it depends very weakly on the cavitation levels. Videos made with a high-speed camera allow an interesting understanding of the fluid flow and the cavitation creation. Moreover, CFD simulations of this experiment, limited to the single phase flow are compared to the experiments. Through the experimental testing, visual observations, and computational predictions, the role of the cavitation on the fluid forces is better understood.

## Introduction

The accurate evaluation of the forces acting on the stem is important for actuators sizing and stem guide selection in order to insure proper functionality and safety. It is recognized that a significant part of the total resultant force come from the flow-induced components. The flow-induced forces acting on the disc can be decomposed into the axial forces (along the stem axis, see Figure 2) and the transverse forces (perpendicular to the stem). For globe valves, the maximum axial load is evaluated analytically taking into account the fluid pressure and the additional sealing force. Additionally, the axial force is measured as part of many standard qualification tests.

In contrast, virtually no publicly available information regarding the transverse force [1 exists (to the knowledge of the authors)]. It exists no relevant data because the transverse force is erroneously judged as insignificant and its measurement requires a sophisticated instrumentation to be placed in the fluid area, thus to be exposed to high pressure and temperature.

Cavitation is a phenomena widely investigated by the scientific community. However, most works focus on noise and the erosion aspect [2][3] and on performance loss due to choking [4][5][6]. No public information is known to the authors on the influence of cavitation on flow-induced force.

An experimental test matrix was developed to evaluate the behavior and dependencies of the axial and transverse forces on factors such as flow rate, cavitation level and valve disc position. An ad-hoc scale 1 valve model was modeled based on an actual 2" Velan globe valve. It includes:
- A Plexiglas body to allow flow visualization
- A heavily instrumented stem to measure forces and moment components

The model was placed in a flow loop allowing for fine-tuning of the flow rate and cavitation conditions.

Additionally, one phase CFD computations were run in order to estimate their capabilities in such domains.

# Experimental setup
## The mock-up
The Plexiglas model body is machined to precisely match the fluid vein of the real two-inch Velan globe valve (see Figure 1).

The stem and the disc (see Figure 2) come from a real Velan valve and are welded together. The stem diameter is reduced from 26.7 mm (1.05 inch) to 14 mm (0.55 inch) over a 90 mm (3.54 inch) section of its length. The stem modifications were made to allow the deflection occurring during the experiment to be measured accurately by the gauges. Twelve strain gauges were fixed to the stem at different positions chosen to allow an accurate measurement of the moment and the resultant forces.

## The loop
The model was placed on the test section of the Modulab loop at the MFEE department at EDF R&D. At this facility, the pump RPM, the temperature and the downstream pressure can easily be set. The downstream pressure can even be sub atmospheric, thus enabling a fine-tuning of the cavitation level.

The fluid media used was common demineralized water and was maintained at 23°C (73 °F) throughout the experiment.

## The test procedure
The experiment was performed under the following operating conditions:
- Valve opening positions: 2, 3, 4, 6 and 16 mm (full open)
- Downstream pressure: 0.4, 0.8, 1.4, 4 bar (absolute)
- Flow rate: 0 to maximum pump capabilities (~70m$^3$/h)

The tests are conducted such that good time-averaged values at each condition set could be obtained. Based on preliminary tests, it is determined that stabilize conditions occur after 30 seconds. The data is measured at 200 Hz.

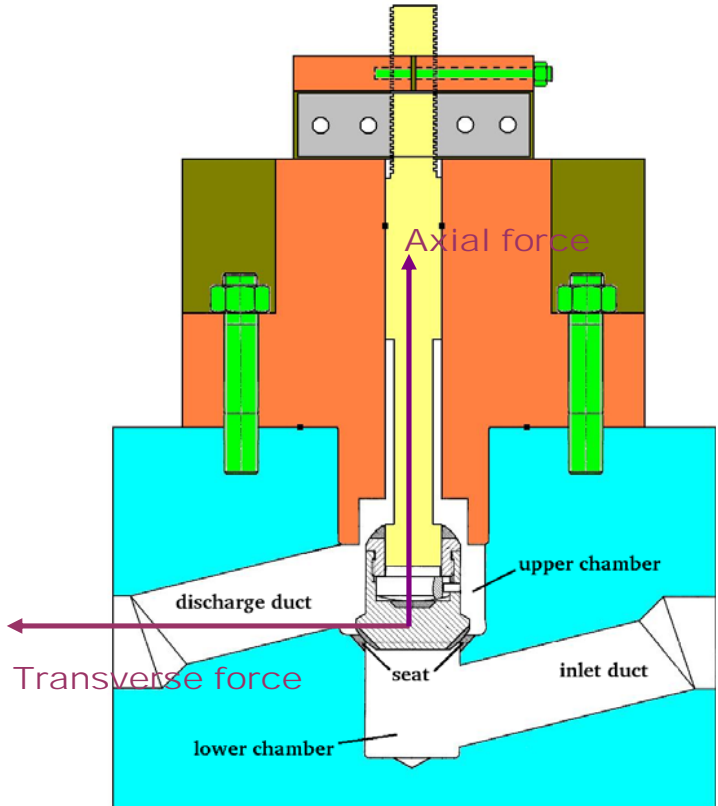

Figure 1. Sketch of the globe valve model.

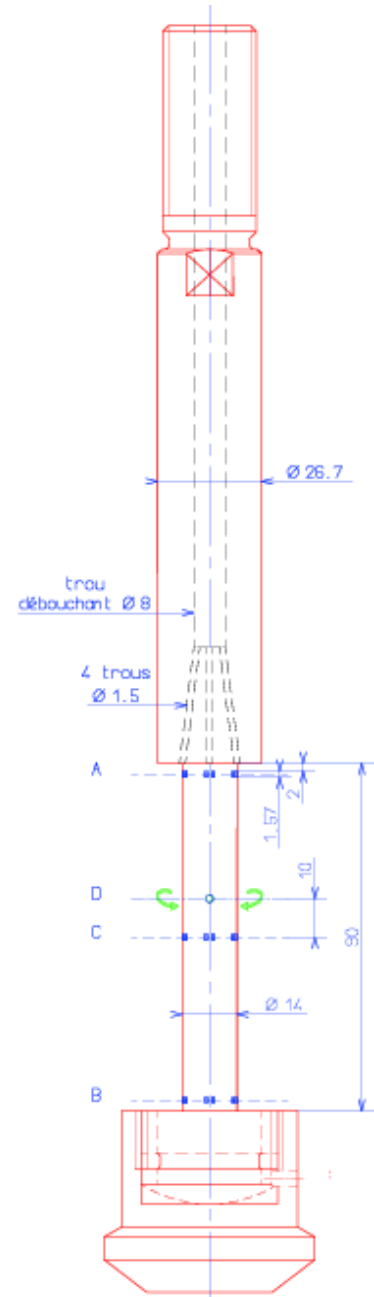

Figure 2. Sketch of the stem and the disc

## Visualization

During the experiment, visual recordings of each test was recorded using a conventional camera. Furthermore, a high-speed camera was used at 13,000 frames per second to record some specific cavitation pattern (see Figure 3 and Figure 4). A more detailed description of the cavitation seen in the mock-up is given in [7] or [8] and some of the movies recorded can be found on the Internet [9].

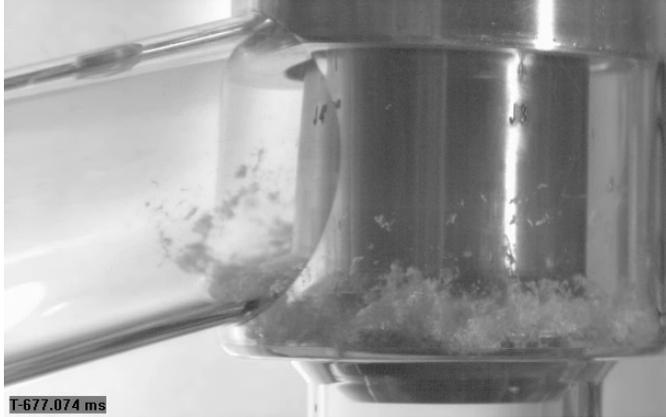

**Figure 3.** opening: 1mm, downstream pressure: 0.8bar, flow rate:15m$^3$/h.

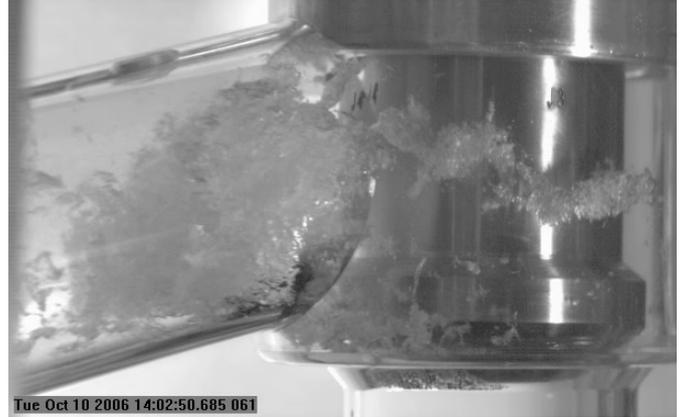

**Figure 4.** opening:4mm, downstream pressure: 0.4bar, flow rate:23m$^3$/h.

# Experimental results
## Spotting cavitation influence
In theory, when the flow is sufficiently turbulent, the flow-induced forces are proportional to the valve pressure drop (DP) and to the flow rate squared (Q²). Using this statement as a basis, it is considered in this paper that any deviation from such a behavior is caused by cavitation. Cavitation, via the well-known phenomena of choking, reduces the valve performance i.e. it decreases the flow rate Q expected at a certain pressure drop DP. Q² and DP are then not proportional to each other anymore.

This gives us two different points of view to study cavitation influence on flow-induced forces. Each of them may lead to different conclusions:
- The DP point of view, which considers that a valve has to undergo a certain pressure drop. This is the point of view favored in this article since it is believed to be more relevant by the author.
- The Q² point of view, which considers the valve at a certain flow rate.

Most of the figures included in this paper show time-average force values using either point of view. On each of these figures, four curves are plotted, one for each specified downstream pressure. Since low system pressure facilitates cavitation creation, cavitation develops at lower DPs and flow rates on low downstream pressure curves than on high downstream pressure curves. Cavitation influence is observed when a non-linear relationship between the measured value and the independent variable occur.

## Cavitation influence on transverse force
### Observations
Figure 5 to Figure 9 show the time-average transverse force versus the pressure drop or the squared flow rate for three disc positions: 2, 6 and 16mm disc positions.

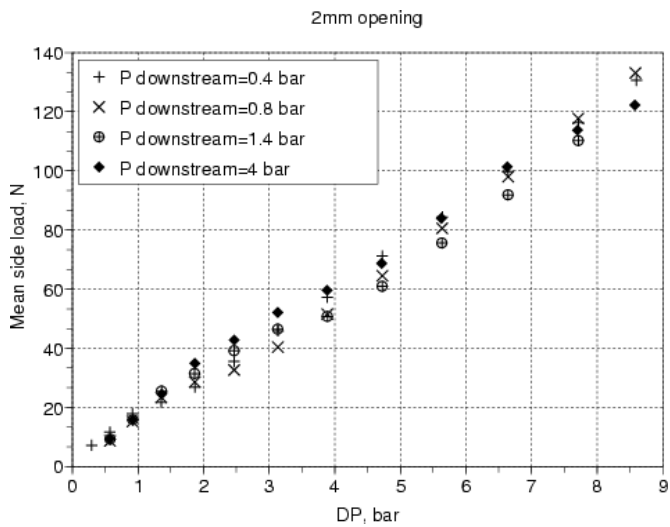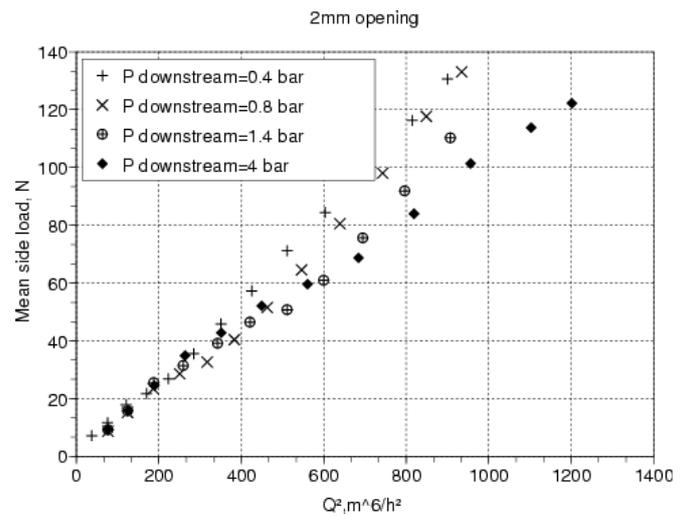

**Figure 5.** Mean transverse force vs. pressure drop at 2mm

**Figure 6.** Mean transverse force vs. squared flow rate at 2mm

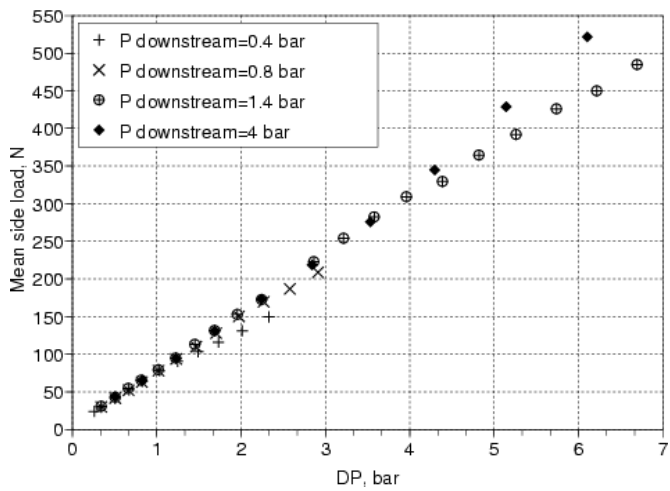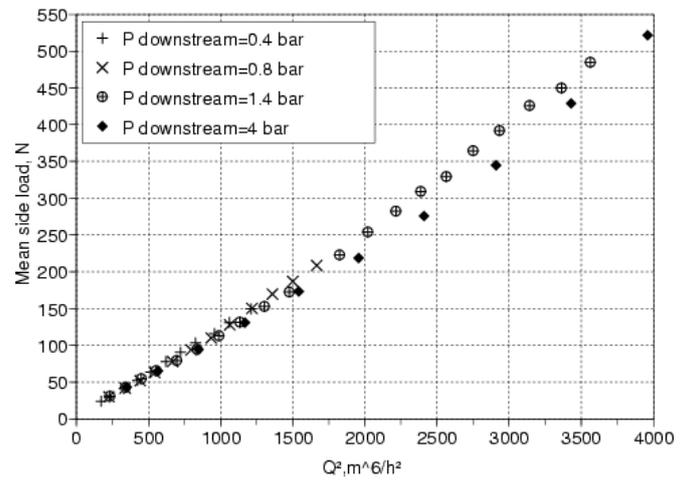

**Figure 7.** Mean transverse force vs. pressure drop at 6mm

**Figure 8.** Mean transverse force vs. squared flow rate at 6mm

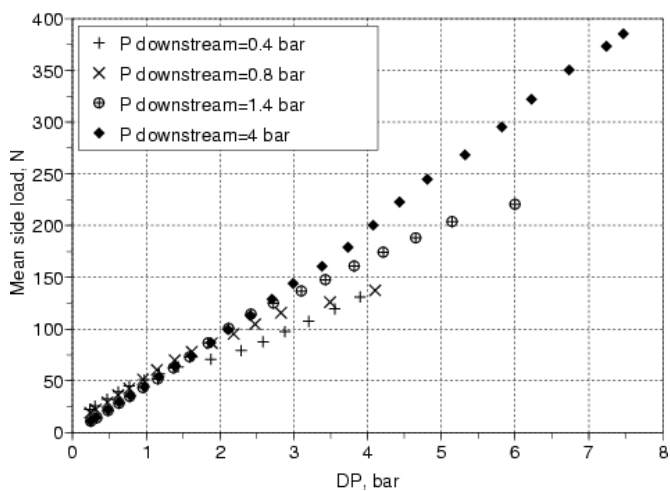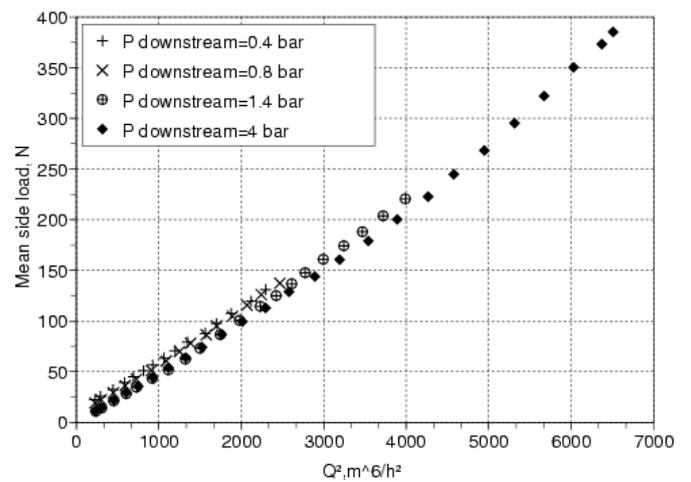

**Figure 9.** Mean transverse force vs. pressure drop at 16mm

**Figure 10.** Mean transverse force vs. squared flow rate at 16mm

At 2mm, the DP point of view (Figure 5) show that the curves remain nearly linear, no significant influence due to cavitation can be observed. From the Q² point of view instead (Figure 6), since choking impedes the flow rate, low downstream pressure (i.e. high cavitation level) curves are shifted to the left. The effects of cavitation are thus seen as an increase of the transverse force. In other words, to overcome the additional pressure drop created by cavitation and to sustain the flow rate, an increase in the upstream pressure (hence DP) is required, consequently the transverse force increases.

At 16mm, the DP point of view illustrates (Figure 9) that as cavitation appears with increasing DP the curves become nonlinear and cavitation reduces the transverse force. From the Q² point of view instead (Figure 10), cavitation has no effect on the transverse force. In other words, the additional pressure drop created by cavitation reduces the pressure drop felt by the stem and thus the transverse force.

At 6mm (Figure 7 and Figure 8), it exhibits a transitional behavior between the two previous observed openings. Other graphs at 3 and 4mm opening (not shown in this paper) confirm the transitional behavior.

Based on the experimental data, the effect of cavitation on the transverse force depends on the valve disc position.

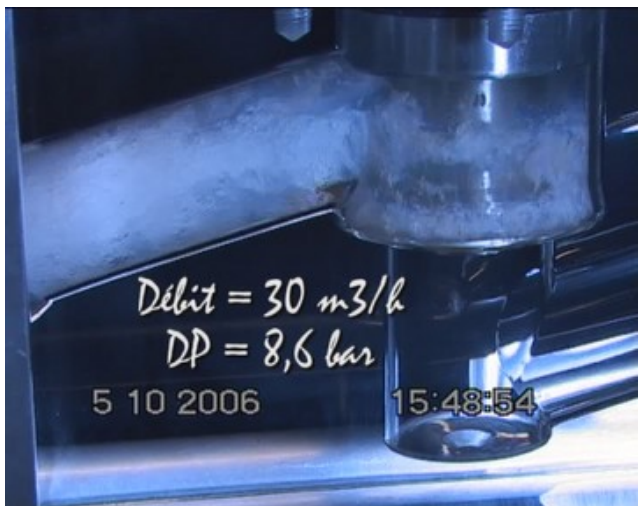

**Figure 11.** cavitation pattern at 2mm opening under high flow rate

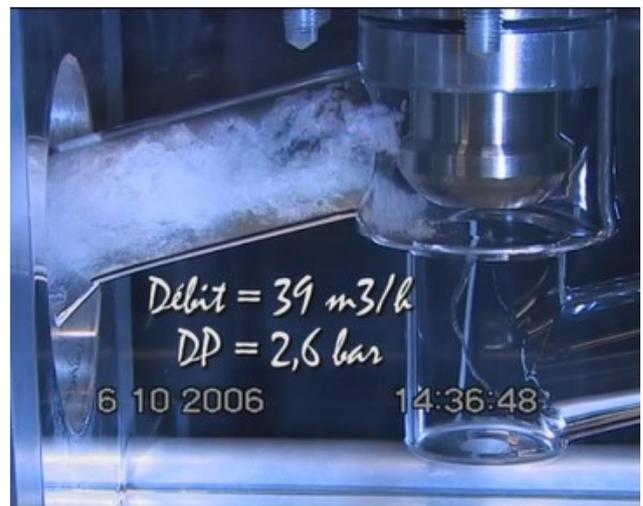

**Figure 12.** cavitation pattern 16mm opening under high flow rate

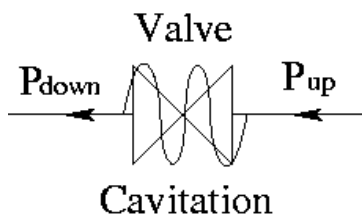

**Figure 13.** scheme of the cavitation presence in the flow at low opening

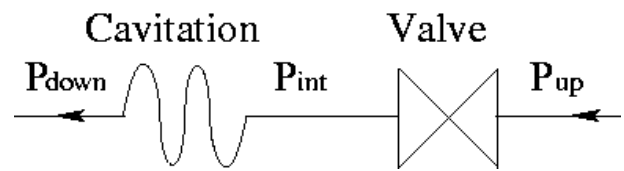

**Figure 14.** scheme of the cavitation presence in the flow at high opening

## Interpretation

Visual observations of the flow inside the mock-up under the highest level of cavitation attained during the experiment show that at small disc positions (Figure 11), the disc head is enveloped by cavitation, whereas at high opening (Figure 12), cavitation is detached and forms downstream from the disc.

Figure 13 and Figure 14 represent schematically the presence of cavitation in the flow. In Figure 13 the disc head is enveloped by caviation, the additional pressure drop created by cavitation is combined with the usual valve pressure drop. In Figure 14 cavitation is detached and the cavitation pressure drop is exerted downstream from the disc. Thus, a theoretical intermediate pressure $P_{int}$ is created between the disc and the cavitation cloud.

According to [6], for a turbulent flow without cavitation, the flow rate is:

$$Q = N_1 * Cv \sqrt{\frac{P_{up} - P_{down}}{\rho/\rho_0}} \qquad (1)$$

and for a turbulent flow with full cavitation:

$$Q = N_1 * C_v * F_l \sqrt{\frac{P_{up} - F_f * P_{vap}}{\rho/\rho_0}} \qquad (2)$$

with:
- $C_v$ : the flow coefficient
- DP : the pressure drop ($P_{up}$ - $P_{down}$)
- $F_f$ : the water critical pressure ratio factor = $0.96 - 0.28\sqrt{P_{vap}/P_c}$ ~ 0.96
- $F_l$ : the pressure recovery factor = 0.79 at 16mm opening
- $N_1$ : a numerical constant dependant on the units used
- $P_c$ : water critical pressure = 220 bar
- $P_{down}$ : pressure far downstream from the valve
- $P_{up}$ : pressure upstream from the valve
- $P_{vap}$ : the vapor pressure = 0.028 bar at 23°C
- $\rho$ : water density
- $\rho_0$ : water density at 15°C

Without cavitation, $P_{int}$ = $P_{down}$. At full cavitation, $P_{int}$ replaces $P_{down}$ in equation (1):

$$Q = N_1 * Cv \sqrt{\frac{P_{up} - P_{int}}{\rho/\rho_0}} \qquad (3)$$

combined with equation (2), it gives:

$$P_{int} = P_{up}(1 - F_l^2) + F_l^2 * F_f * P_{vap} \qquad (4)$$

Considering the thermodynamic conditions of the experiment and the $F_l$ found at 16 mm opening, it yields:

$$P_{int} = 0.3759 P_{up} + 0.017 \qquad (5)$$

Figure 15 shows the transverse force plotted twice versus $P_{up}$-$P_{int}$. Once in black, where $P_{int}$ = $P_{down}$, like in the previous chapter *Observations*, it thus redraws exactly Figure 9.

And once in blue, using equation (5) to evaluate $P_{int}$. Figure 16 is just a close-up of Figure 15. From Figure 16 it can be seen that as cavitation appears with increasing $P_{up}$-$P_{int}$, the black curves detach one by one from the linear behavior whereas the blue curves reattach one by one. It passes from the validity range of Equation (1) to the validity range of Equation (2).

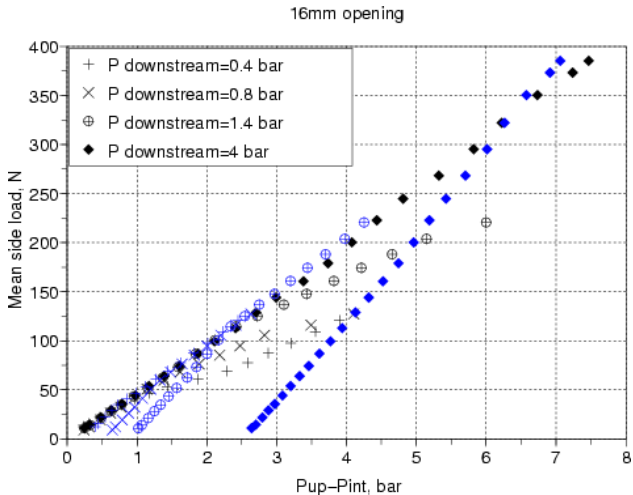 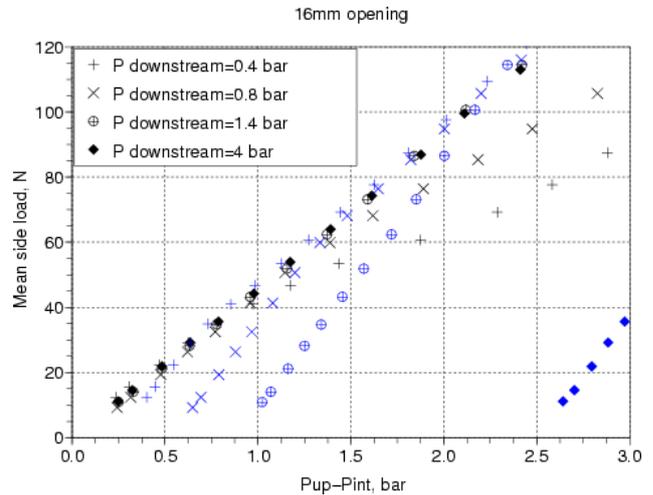

**Figure 15.** Mean transverse force vs. $P_{int}$. In black: $P_{int}$= $Pd_{own}$, in blue: $P_{int}$=0.3759$P_{up}$+0.017.

**Figure 16.** Zoom from Figure 15. Mean transverse force vs. $P_{int}$. In black: $P_{int}$= $Pd_{own}$, in blue: $P_{int}$=0.3759$P_{up}$+0.017.

In conclusion, at large disc positions ( i.e. when cavitation is located far from the disc) the transverse force has a linear behavior versus $P_{up}$-$P_{int}$ and $P_{int}$ can be estimated using Equation 4.

## Conclusion

Cavitation does not cause any dramatic raise of transverse force, whatever is the point of view used. Its influence on transverse force depends on the opening of the valve. An interpretation based on cavitation location is proposed.

From the DP point of view, cavitation acts only as a transverse force limiter at large disc openings. From the $Q^2$ point of view, cavitation increases the transverse force at small disc positions. However, since much lower values have been observed at low opening, it seems reasonable to conclude that cavitation act as a general limiter.

## Axial force

Figure 17 to Figure 20 show the average axial load versus DP at various opening positions. At 2mm opening (Figure 17), the stem is compressed linearly as DP increases, but as the valve opens, the acceleration of the fluid under the disc generates low pressures and thus a traction force. At 6mm (Figure 19), traction and compression cancel each other and at 16 (Figure 20), their addition results in a traction. From Figure 20, it is seen that cavitation reduces the final traction force. No other important effect on axial load has been recorded.

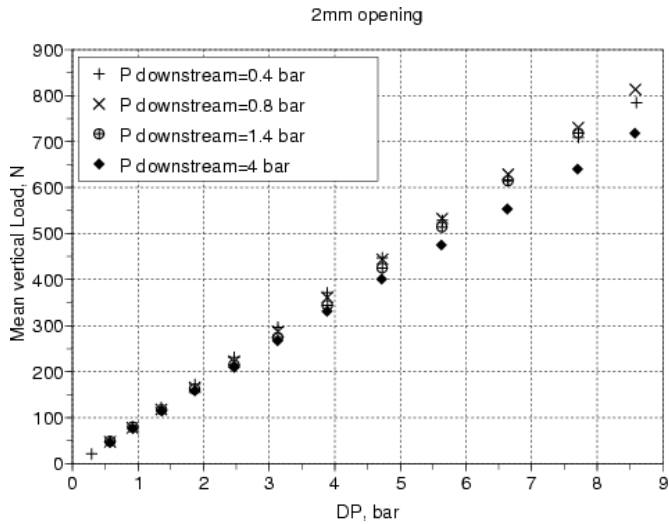
**Figure 17.** Mean axial load vs. pressure drop at 2mm

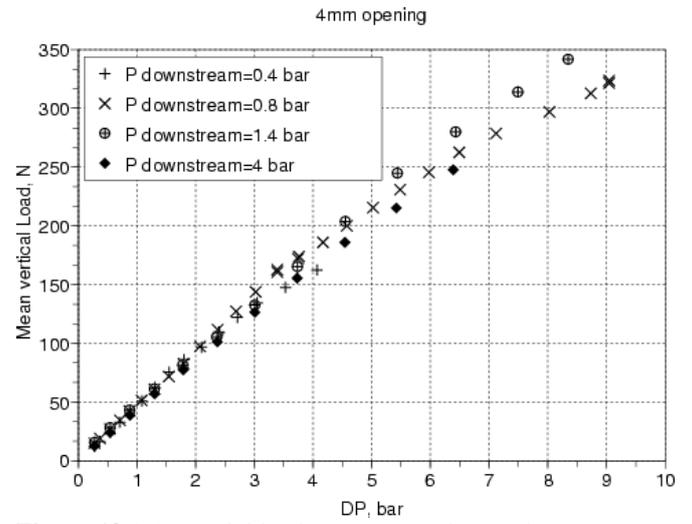
**Figure 18.** Mean axial load vs. pressure drop at 4mm

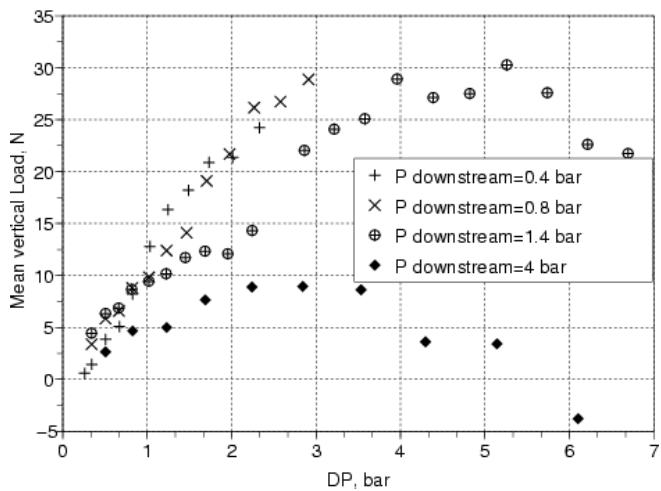
**Figure 19.** Mean axial load vs. pressure drop at 6mm

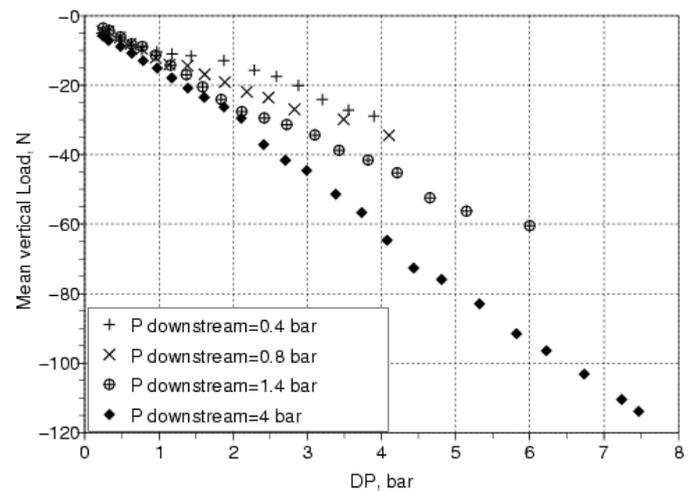
**Figure 20.** Mean axial load vs. pressure drop at 16mm

## Maximum transverse force position

At low opening position, the flow around the disc is almost axisymmetric resulting in nearly no transverse force. At large disc position, the disc is retracted out of the main flow area and is thus less exposed, which reduces the transverse force values. The maximum transverse force is consequently at a mid stroke position.

To evaluate this position, the hydraulic surface is used :

$$\text{Hydraulic surface} = \frac{\text{Average transverse force}}{\text{Pressure drop}} \qquad (6)$$

It is computed using data recorded without cavitation.
Figure 21 shows the hydraulic surface found from the experiment and the simulation. The experimental results increase until 6mm and then show a lower value at 16mm opening. Additional data between those two points would be needed to accurately determine the maximum transverse force position. Numerical results show a maximum at 8mm.

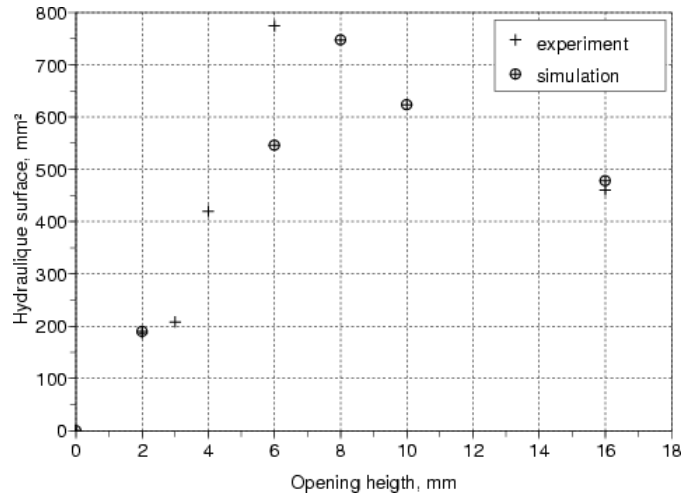

**Figure 21.** Hydraulic surface vs. Disc position

## Transverse force vs. axial force

The importance of transverse force can only be estimated against the scale of axial force. In this chapter, as an example, in-plant extreme conditions are considered:
- upstream pressure    $P_{up}$ = 270bar
- downstream pressure  $P_{down}$ = 20 bar

The maximum axial force occurs when the valve is closed. The analytical formula below takes into account the fluid force and the seating force (contact force between the seat and the disc to insure sealing):

$$F_{Axial,\max} = \frac{\pi D_{seat}^2}{4}(P_{up} - P_{down}) + \pi D_{stem}^2 P_{down} + E\pi D_{seat} \qquad (7)$$

$D_{seat}$ and $D_{stem}$ are the seat (45 mm) and stem (26.9 mm) diameters and E is the linear contact pressure: for a sphere-cone sealing, E = 150N/mm is stipulated. This gives:

$$F_{Axial,\max} = 46200 N \qquad (8)$$

The maximum transverse force is estimated using the maximum hydraulic surface experimentally found (see chapter *Maximum transverse force position*) : 774 mm² at 6mm, cavitation is thus not taken into account. It gives:

$$F_{Tran,\max} = S_{\max} * (P_{up} - P_{down}) = 11600 N = 25\% * F_{Axial,\max} \qquad (9)$$

Therefore, during its life-time, such a globe valve is subjected to transverse force of the order of 25% the axial force for which it was sized, which could lead to wear or galling of the stem and guide.

## Numerical results

The computational Fluid Dynamics (CFD) software package ANSYS CFX is used to predict the resultant flow field and flow-induced forces at 2, 6, 8, 10, 16 mm. Three different simulations are performed at each disc position. The nominal valve differential pressures used for the three simulations are 0.5, 1 and 2 bar.

The computational domain includes the fluid enclosure of the body, 20 pipe-diameters of upstream pipe, and 31 pipe-diameters of downstream pipe. The flow is assumed to be fully turbulent and the turbulence is modeled using the k-ε RNG turbulence model. Additional runs are performed using the k-ω SST model and good agreement between the two turbulence models exists.

## Boundary Conditions

The boundary conditions used in the CFD analyses consisted of an inlet velocity boundary, a static pressure outlet, and no-slip frictional walls for the pipe walls, internal body surfaces, and disc. The inlet velocity profile was defined using the 1/7th power-law profile given by Eq. (5) below. The maximum velocity was chosen to produce the desired flow rate through the valve.

$$V(r) = V_{max}(1 - r/R)^{1/7} \qquad (5)$$

The turbulence inflow condition at the inlet was defined by setting the turbulence intensity to 5%. The static pressure was defined at the outlet and was set to 4 bar (absolute).

## Numerical Results

A comparison of the hydraulic surface for the axial and transverse force (shown in Figure 21 and Figure 22 provide a comparison of experimental and numerical results. In general, the experimental and numerical results are in good agreement; however, the CFD prediction of the transverse force at 6 mm is about 30% less than the experimental data.

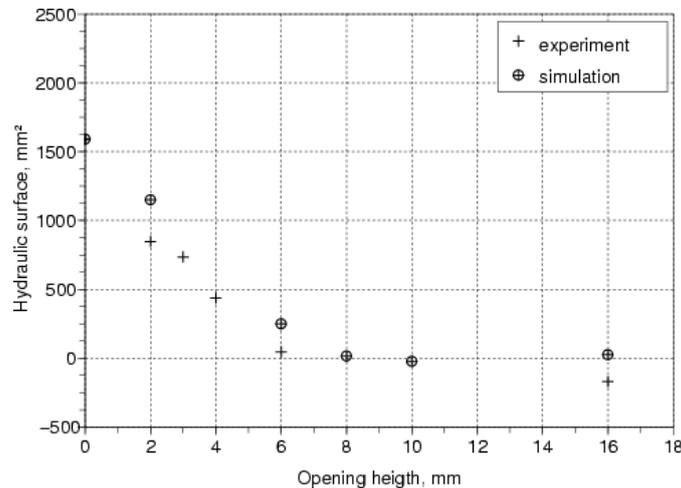

Figure 22 The hydraulic area based on experimental data and numerical results for the axial force are compared.

In addition to the experimental observations, the CFD predictions provide additional insight into the flow field. The pressure contours on the disc and along the internal body surfaces help illustrate the relationship between the axial and transverse hydraulic surface based on disc position.

## Disc and Internal Body Surface Pressure Contours

The axial and transverse forces are better understood by examining the pressure contours on the disc head. The axial load is strongly dependent on the pressure differences between the upstream and downstream disc surfaces that are perpendicular to the disc axis (e.g., the disc nose). The transverse force is dependent on the pressure contours on the sides of the disc. In Figure 23 the pressure contours are shown on the disc and internal body surface. At small disc positions the upstream static pressure is nearly equal to the stagnation pressure causing a large pressure force acting along the disc axis. As the disc position increases and the flow rate increases the static pressure in the upstream duct decreases from the stagnation pressure and increases the transverse force.

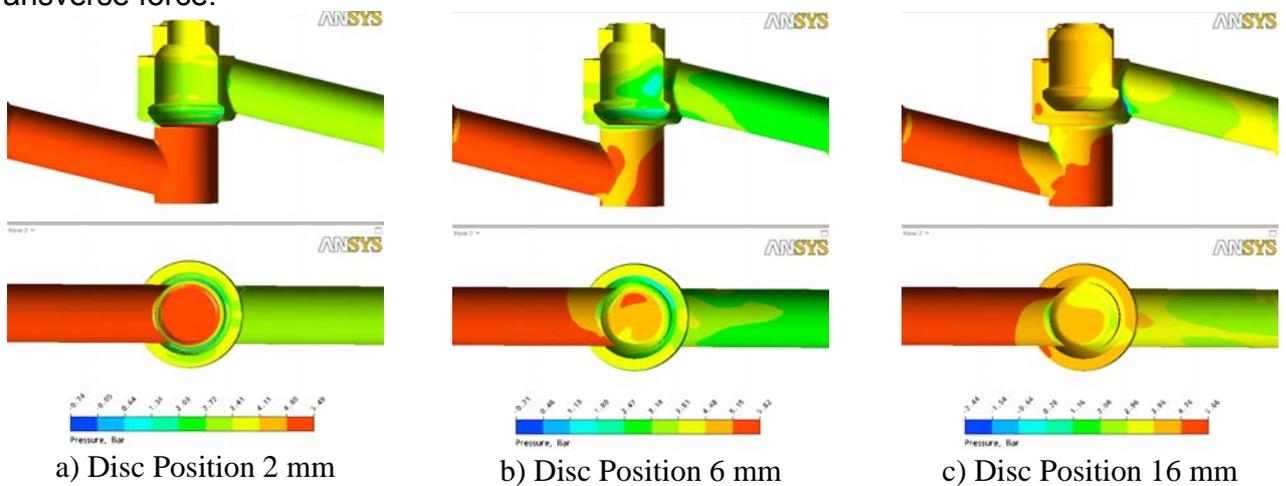

a) Disc Position 2 mm     b) Disc Position 6 mm     c) Disc Position 16 mm

**Figure 23 The pressure contours on the disc and internal body surface are shown at three disc positions : (a) 2 mm (b) 6 mm, and (c) 16 mm (flow is from left to right). The axial load is primarily governed by the pressure acting on the disc nose (bottom view) and the transverse force is governed by the pressure acting from left to right (top veiw).**

## Streamlines

The streamlines through the valve body are shown for three disc positions 2 mm, 6mm, and 16 mm. The streamlines are colored by velocity magnitude. At small disc positions the highest velocities occur in the narrow seat gap. At larger disc position the flow velocity throughout the domain is increased.

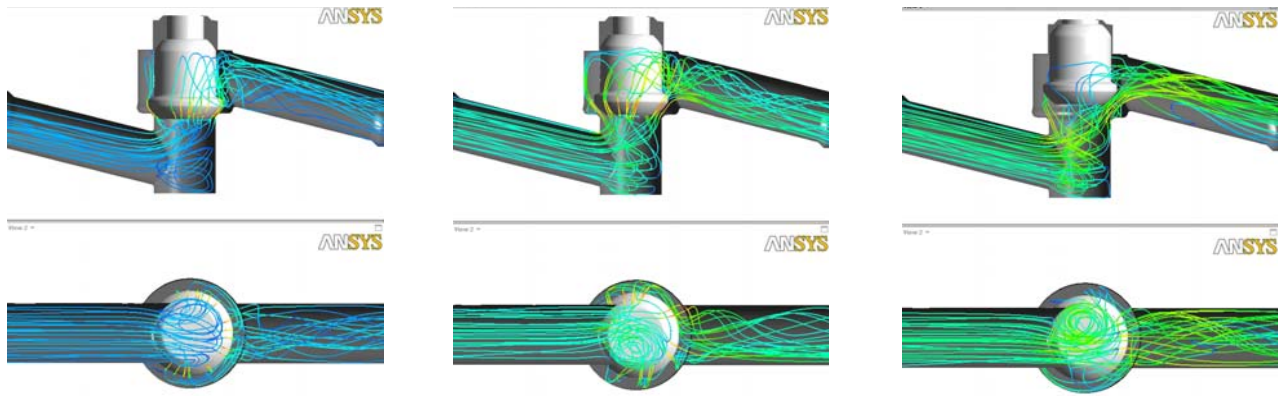

| a) Disc Position 2 mm | b) Disc Position 6 mm | c) Disc Position 16 mm |

**Figure 24 The streamlines in the fluid enclosure (flow is from left to right) are shown at three disc positions : (a) 2 mm (b) 6 mm, and (c) 16 mm. The axial load is primarily governed by the pressure acting on the disc nose (bottom view) and the transverse force is governed by the pressure acting from left to right (top veiw).**

# Conclusion

A 2" Plexiglas globe valve model designed to permit visual observation and measurements of flow-induced force on the stem was built. An extensive test program was conducted with varying key parameters such as disc position and operating conditions.

No dramatic increase of force from cavitation has been observed. Instead, it appeared to act as a force limiter at high disc opening by reducing the pressure drop seen from the disc.

The maximum transverse force was observed at intermediate opening (between 6 and 10mm) and it was calculated that it could reach 25% of the maximum axial force.

Numerical predictions were in good agreement with experimental data and together with the analysis of experimental results supplied a better understanding of the flow-induced force on the stem.

# Acknowledgements

A.Archer, D.Hersant, JF.Rit and the technical teams from EDF R&D are acknowledged for their fruitful contributions and their enthusiasm.

# References

[1]     Hosler J., "Evaluation of Globe Valve Side-Loading Using Computational Fluid Dynamics Modeling", EPRI report 1006649, 2001.
[2]     Brennen C.E. "Cavitation and Bubble Dynamics", ISBN 0-19-509409-3, 1995.
[3]     Young F.R., "Cavitation", ISBN: 978-00-7707-094-6, 1989.
[4]     Tullis J.P., "Hydraulics of Pipelines, Pumps, Valves, Cavitation, Transients", ISBN 0-471-83285-5, 1989.
[5]     Miller D.S. "Internal Flow Systems", ISBN: 0-947711-77-5, 1990.
[6]     AFNOR NF "Industrial-process control valves. Part 2-3: flow capacity. Test procedures.", EN 60534-2-3, 1998.
[7]     Ferrari J. & Leutwyler Z., "Fluid flow force measurement under various cavitation state on a globe valve model", Proceedings of 2008 Pressure and Vessels Piping ASME conference, july 27-30, 2008, Chicago, Illinois, USA.
[8]     Ferrari J."Cavitation et efforts dans un robinet à soupape de DN 50, observations sur une maquette à l'échelle 1. " technical report H-T21-2007-00093-FR, 2007, freely available on demand.
[9]     http://www.dailymotion.com/jerome_ferrari